\documentclass[twocolumn]{aastex63}
\usepackage{rotating}
\usepackage{graphicx}
\usepackage{amssymb}
\usepackage{amsmath}
\usepackage{hyperref}
\usepackage{lineno}


\def\gtrsim{\mathrel{\hbox{\rlap{\hbox{\lower4pt\hbox{$\sim$}}}\hbox{$>$}}}}
\def\lesssim{\mathrel{\hbox{\rlap{\hbox{\lower4pt\hbox{$\sim$}}}\hbox{$<$}}}}
\def\farcs{\hbox{$.\!\!^{\prime\prime}$}}

\begin{document}

\title{XMM-Newton and Chandra observations of the candidate Fermi-LAT pulsar 4FGL J1015.5-6030}
\author{Jeremy Hare}
\altaffiliation{NASA Postdoctoral Program Fellow}
\affiliation{NASA Goddard Space Flight Center, Greenbelt,  MD 20771, USA}
\author{Oleg Kargaltsev}
\affiliation{Department of Physics, The George Washington University, 725 21st St. NW, Washington, DC 20052}
\affiliation{The George Washington Astronomy, Physics, and Statistics Institute of Sciences (APSIS)}
\author{George Younes}
\affiliation{NASA Goddard Space Flight Center, Greenbelt,  MD 20771, USA}
\affiliation{Department of Physics, The George Washington University, 725 21st St. NW, Washington, DC 20052}
\affiliation{The George Washington Astronomy, Physics, and Statistics Institute of Sciences (APSIS)}
\author{George G. Pavlov}
\affiliation{Department of Astronomy \& Astrophysics, Pennsylvania State University, 525 Davey Lab, University Park, PA 16802, USA}
\author{Igor Volkov}
\affiliation{Department of Physics, The George Washington University, 725 21st St. NW, Washington, DC 20052}

\email{jeremy.hare@nasa.gov}

\begin{abstract}
4FGL J1015.5-6030 is an unidentified Fermi-LAT source hosting  a bright, extended X-ray source whose X-ray spectrum is consistent with that of a young pulsar, yet no pulsations have been found. Here we report on XMM-Newton timing and Chandra imaging observations of the  X-ray counterpart of 4FGL J1015.5-6030. We find no significant periodicity from the source and place a 3$\sigma$ upper-limit on its pulsed fraction of 34$\%$. The Chandra observations resolve the point source from the extended emission. We find that the point source's spectrum is well fit by a blackbody model, with temperature $kT=0.205\pm0.009$ keV, plus a weak power-law component, which is consistent with a thermally emitting neutron star with a magnetospheric component. The  extended emission spans angular scales of a few arcseconds up to about 30$''$ from the point source and its spectrum is well fit by a power-law model with a photon index $\Gamma=1.70\pm0.05$. The extended emission's spectrum and 0.5-10 keV luminosity of 4$\times10^{32}$ erg s$^{-1}$ (at a plausible distance of 2 kpc) are consistent with that of a pulsar wind nebula. Based on a comparison to other GeV and X-ray pulsars, we find that this putative pulsar is likely a middle-aged (i.e., $\tau\sim 0.1$--1 Myr) radio-quiet pulsar with $\dot{E}\sim10^{34}-10^{35}$ erg s$^{-1}$.
\end{abstract}

\section{Introduction}
During its $>$ 14 years of operation the Fermi Large Area Telescope (Fermi-LAT) has uncovered many new Galactic  GeV sources.  However, a large fraction of them ($\sim60\%$) still remain unidentified. A typical strategy to identify these GeV sources is to search for a counterpart at lower frequencies. Many GeV emission mechanisms (e.g., curvature radiation, inverse Compton scattering) require an energetic population of particles that are often detected via their synchrotron emission at X-ray energies (see e.g., \citealt{2016ApJ...833..143L,2019ApJ...887...18K,2021AJ....161..154K}).

4FGL J1015.5-6030 (J1015 hereafter; formerly 3FGL J1016.5-6034) is an unidentified Fermi-LAT source that was recently found to have a relatively bright X-ray counterpart near the center of the GeV source positional uncertainty ellipse \citep{2019ApJ...875..107H}. This X-ray counterpart was resolved into a point source surrounded by extended emission on both small ($\sim30''$) and large ($\sim5'$) scales. The source's spectrum was well fit by a blackbody (BB) plus power-law (PL) model. No longer wavelength counterpart (i.e., radio, IR, optical) to the source was found.  The source is most likely to be a pulsar with a pulsar wind nebula (PWN) based on its X-ray spectrum, the point-like plus extended X-ray emission, and the fact that it's a GeV source. However, the previous X-ray observation with the Chandra X-ray observatory (CXO) was very short (only 2 ks to obtain an accurate source position) and did not provide enough photons to characterize the point source and the extended emission. Additionally, the XMM-Newton observation was performed in Full Frame mode, lacking the time resolution to search for pulsations with frequencies higher than $\sim 7$ Hz, and no pulsations were found.

Here we report the results of the latest X-ray observing campaign of J1015 with CXO and XMM-Newton. In Section \ref{obs_and_dat} we discuss the observations and data reduction. In Section \ref{results} we report the results of our analysis, while in Section \ref{discuss} we compare these results to the broader population of GeV and X-ray pulsars. We summarize our findings and conclusions in Section \ref{sumandconc}.

\section{Observations and Data Reduction}
\label{obs_and_dat}

\subsection{XMM-Newton}
\begin{figure*}[t!]
\centering
\includegraphics[trim={0 0 0 0},scale=0.42]{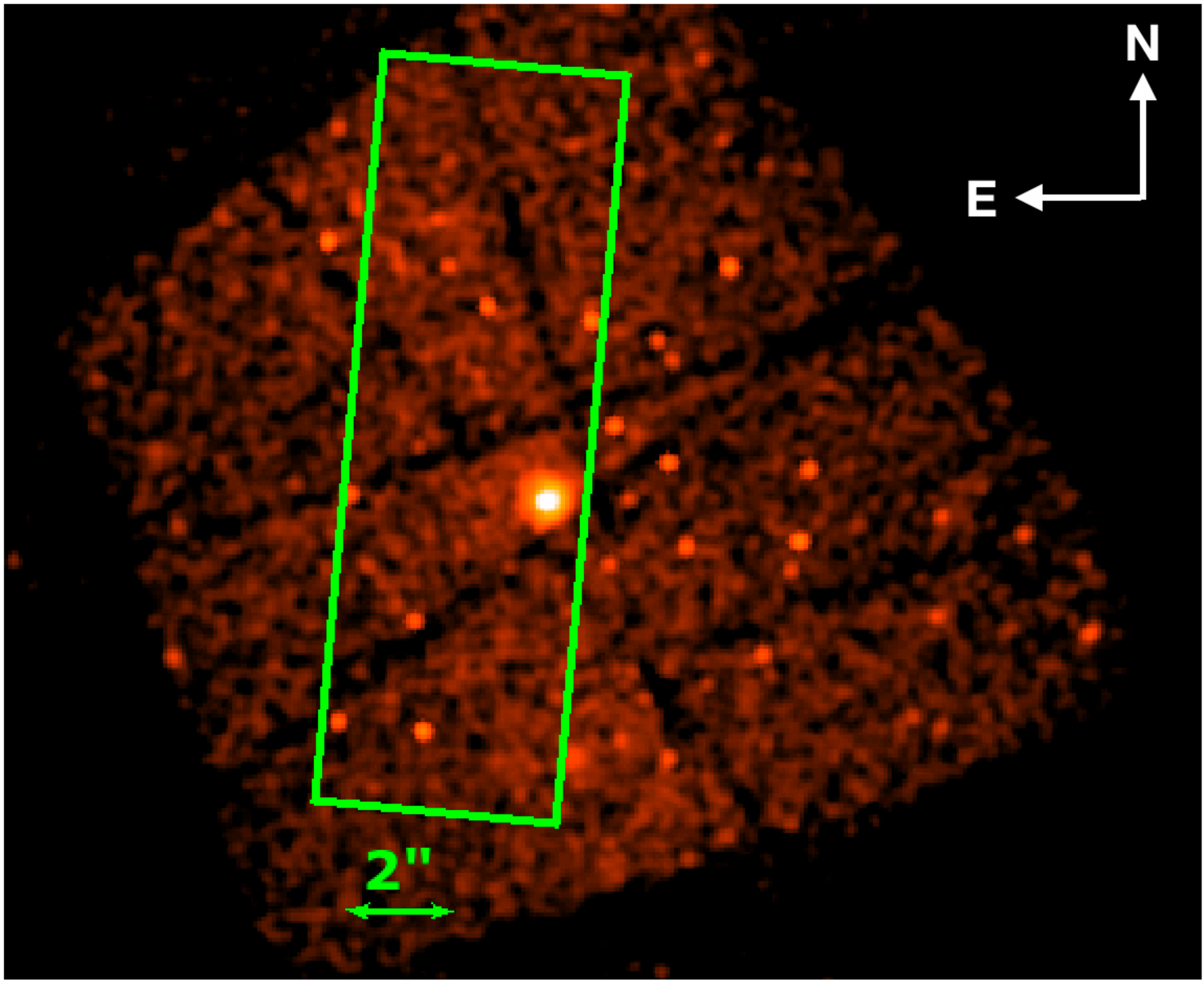}
\includegraphics[trim={0 0 0 0},scale=0.415]{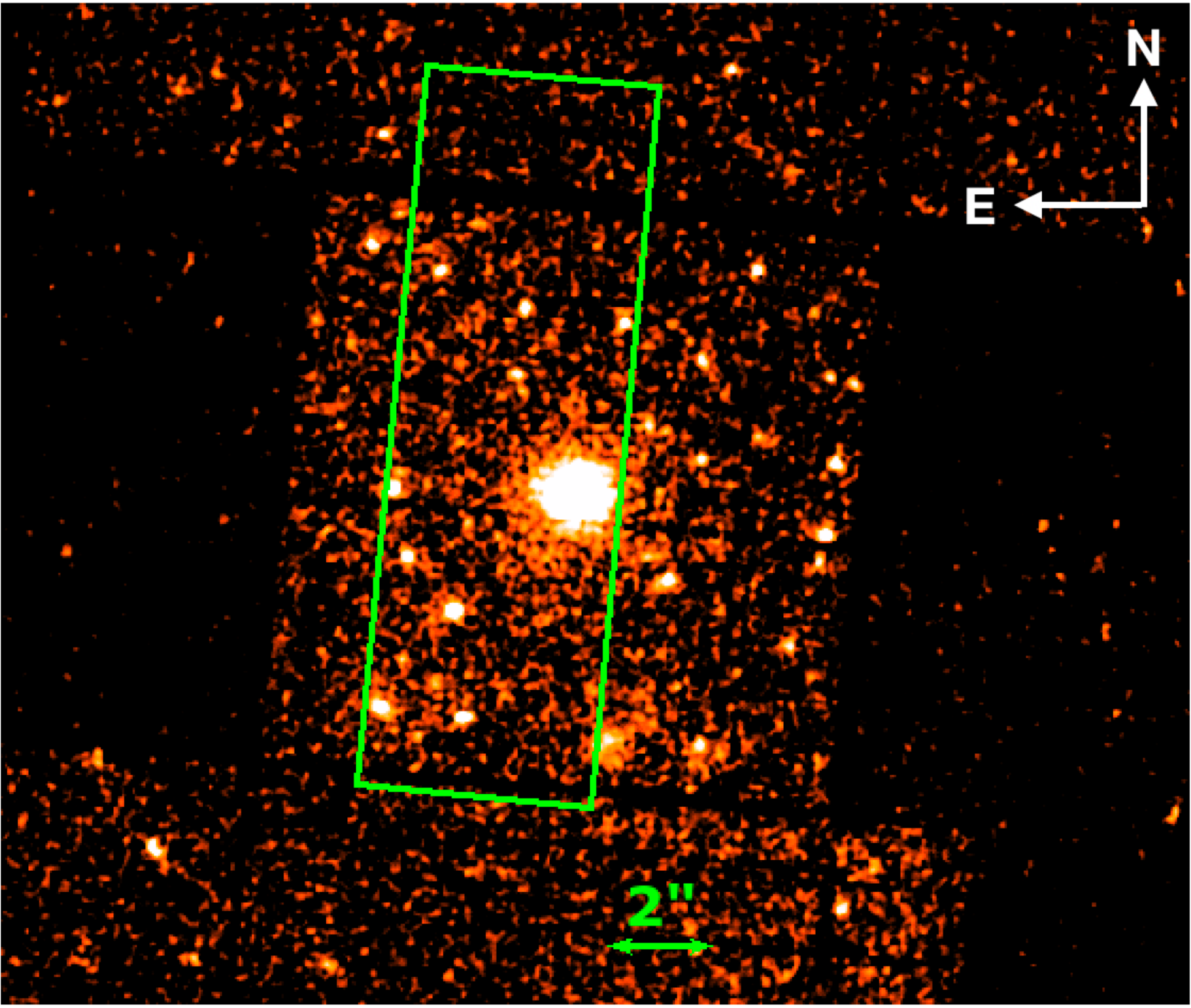}
\caption{{\sl Left:} Merged 0.5-8 keV CXO image binned by a factor of 8 and smoothed with a $r=3''$ Gaussian kernel. The thick green box shows the size and pointing of the XMM-Newton pn timing mode observation. {\sl Right:} Merged 0.5-10 keV XMM-Newton MOS1+MOS2 from the latest observation smoothed with a $r=3''$ Gaussian kernel.  The thick green box shows the size and pointing of the XMM-Newton pn timing mode observation. 
\label{timing_mode}
}
\end{figure*}

XMM-Newton observed J1015 on 2020 February 15 (ObsID 0853180101) for approximately 51 ks. The EPIC pn detector was operated in Timing mode, while MOS1 and MOS2 were operated in Full Frame mode (see Figure \ref{timing_mode} for the placement of the Timing mode observation on the source region). The data were reduced and analyzed using version 20.0.0 of the  Science Analysis System (SAS). The pn Timing mode and MOS Full Frame mode event lists were cleaned, including the removal of times with high particle background, and reduced following standard SAS procedures. The MOS data were relatively unaffected by background particle flares, leaving about 50.7 ks of exposure time after filtering. The PN timing mode data were also mostly unaffected by background particle flares, however, there were two relatively strong soft flares\footnote{See \url{https://xmm-tools.cosmos.esa.int/external/xmm_user_support/documentation/sas_usg/USG/sfevtpntiming.html}}. After removing these flares $\approx46.8$ ks of exposure time remained, while the observation spanned a total of about 49.4 ks. All event arrival times were corrected to the solar system barycenter before performing timing analysis using SAS's {\tt barycen} task. 

We also reanalyzed the data from our previous XMM-Newton observation (ObsID 080293010; see Section \ref{spec_ext}) reported in \cite{2019ApJ...875..107H}. The data were reprocessed and cleaned following the standard procedures and using version 20.0.0 of SAS to minimize the calibration differences between the new and old data sets (the old data set was previously reduced with an older version of SAS). Scientific exposures of 16.4, 16.4, and 12.0, remained for MOS1, MOS2, and PN, respectively, after cleaning.

\subsection{{\sl Chandra} X-ray Observatory}

The CXO observed J1015 with the Advanced CCD Imaging Spectrometer (ACIS; \citealt{2003SPIE.4851...28G}) on two separate occasions. The first observation (ObsID 22455) took place on 2021 March 28 having an exposure time of 29.73 ks, while the second (ObsID 24777) took place on 2021 September 11, having an exposure time of 28.72 ks. For both observations, J1015 was imaged on the ACIS-I detector, which was operated in timed exposure mode using the Very Faint telemetry format. This detector setup provides a time resolution of 3.2 s. Both observations were reprocessed and analyzed using the Chandra Interactive Analysis of Observations (CIAO;  \citealt{2006SPIE.6270E..60F}) package version 4.13 and the 4.9.6 version of the calibration database.

\subsubsection{Source Position}

The previous $\sim 2$ ks CXO observation was too short to detect enough additional sources to correct CXO's absolute astrometry \citep{2019ApJ...875..107H}. However, the longer 60 ks observation has allowed us to accomplish this and to retrieve a more accurate absolute position of the source. To achieve this, we first ran {\tt wavdetect} on both observations, and then merged the observations to one another with the CIAO tools  {\tt wcs\_match} and {\tt wcs\_update} using the X-ray source positions and ObsID 24777 as the reference image. Then we ran {\tt wavdetect} on the merged image to detect fainter sources than in the individual images. We then cross-matched the on-axis ($<5'$ from the pointing) X-ray sources detected with a significance  $>5$  to Gaia DR3 \citep{2016A&A...595A...1G,2021A&A...649A...1G} sources using a 1$''$ search radius and  excluding the X-ray counterpart to J1015. We found eight sources that matched and we used these to correct the absolute astrometry using {\tt wcs\_match} and {\tt wcs\_update}. The root-mean-square (RMS) residual offsets between sources were $0\farcs{51}$ prior to the correction, which improved to $0\farcs{25}$ after the correction was applied. We adopt this value as the 1$\sigma$ uncertainty on the absolute astrometry and find that it dominates the uncertainty in the source position as the source is relatively bright, leading to small statistical uncertainties. After correcting the astrometry we find that an updated source position of R.A.$=153\fdg94156$, decl.$=-60\fdg494389$ with a 1$\sigma$ positional uncertainty of $0\farcs{25}$ that we conservatively round up to $0\farcs{3}$ and which includes both statistical and astrometric uncertainties added in quadrature.

\begin{figure*}[t!]
\centering
\includegraphics[trim={0 0 0 0},scale=0.55]{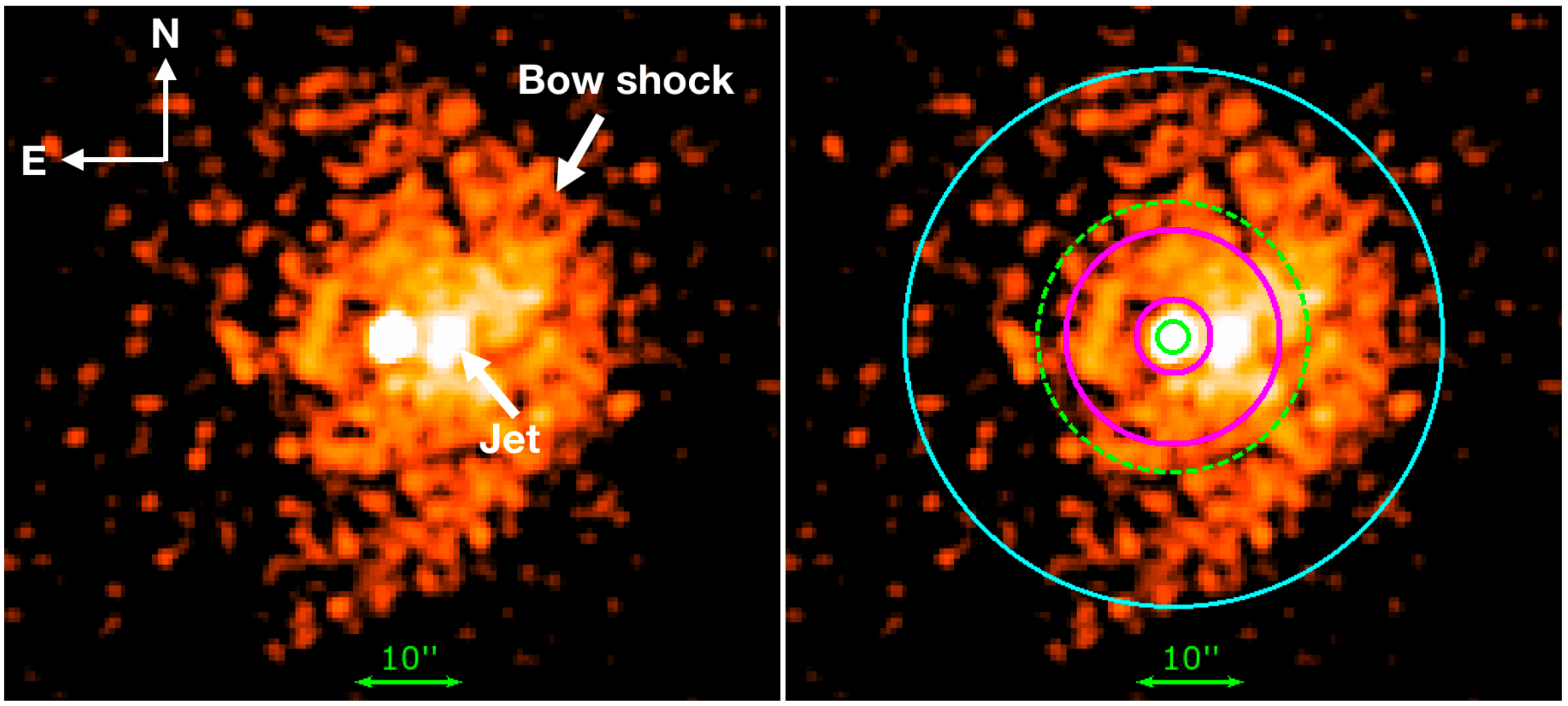}
\caption{{\sl Left:} CXO ACIS 0.5-8 keV summed images of the pulsar and PWN. The image is smoothed with an $r=3''$ Gaussian kernel. Several features of the PWN are shown by the white arrows and corresponding labels. {\sl Right:} The same image but with the regions used for spectral extraction over-plotted. The green inner circle was used for the point source, with the magenta annulus used for the background. The same magenta annulus was used for the inner PWN, with the region between the green dashed circle to the cyan circle used for the background. The  region between the green dashed circle to the cyan circle was used for the outer PWN. For the Full PWN, we used the region between the inner radius of the magenta annulus to the cyan circle. The background regions for the full PWN and outer PWN are not shown (see Section \ref{spec_ext}).
\label{merged_obs}
}
\end{figure*}

\subsubsection{Image analysis}
\label{img_analysis}

To increase the statistics and quality of the image, we merged the two CXO observations together using the CIAO tool {\tt merge\_obs}. In the merged 0.5-8 keV image, the bright point source and surrounding extended diffuse emission, 
originally reported by \cite{2019ApJ...875..107H}, are clearly visible (see Figure \ref{merged_obs}). The extended emission appears to consist of at least two components, the first is the brighter inner component on a scale of $\sim5''-10''$  and the second is the fainter, but larger, component on a scale of $\sim10''-30''$.
The small scale PWN shows additional features, such as a bright spot with surrounding extended emission to the west of the point source, and a ring like feature to the east of the point source. The larger-scale PWN shows a possible bow shock-like morphology, being rounded on the western side and flat on the eastern side (see Figure \ref{merged_obs}).

The two CXO observations were taken roughly six months apart. This has afforded us the opportunity to search for variability in the PWN morphology and brightness, as has been observed in other young pulsars (see e.g., \citealt{2001ApJ...554L.189P,2006ApJ...640..929D,2013ApJ...763...72D,2018ApJ...868..119K}). Indeed, there does appear to be differences in the inner PWN morphology and brightness observed between these two observations. The exposure corrected images, zoomed in on the inner PWN, are shown in Figure \ref{expo_innerPWN}. The radial surface brightness profile around the point source shows 
marginal evidence ($\approx2.2\sigma$) for a change in the inner PWN brightness between observations\footnote{Note that the profiles are not corrected for exposure time, but that the difference in exposure times between the two observations is $<4\%$.} at a distance of $\sim$7$''$ from the point source (see Figure \ref{radial_prof}).

\subsubsection{Spectral extraction}
\label{spec_ext}
 We extracted the CXO spectra for the point source and extended emission from each observation separately using the CIAO tool {\tt specextract}. The spectra of the point source were extracted from an $r=1\farcs{5}$ circular region centered on the point source position. For the extended emission, we used several different regions to look for changes in the spectrum of the PWN as a function of distance from the point source. First, we extracted spectra from the full PWN using an annulus $3\farcs5<r<25''$ centered on the point source. The background region for these spectra was chosen as a source-free region offset from the PWN.  Next, we extracted the spectra of the smaller-scale PWN from an annulus $3\farcs{5}<r<10''$. For these spectra, the background was taken from an  $12\farcs6<r<25''$ annulus containing the larger PWN (see Figure \ref{merged_obs}). Lastly, we extracted the spectrum of the outer PWN from an $12\farcs6<r<25''$ annulus centered on the point source, and used a source-free region offset from the PWN for the background.  This source-free region was also used as the background spectrum for the point source. Prior to performing spectral analysis, all CXO spectra were binned to have at least one count per bin so that  W-statistics (which is a variant of Cash statistics; \citealt{1979ApJ...228..939C}) could be used\footnote{See \url{https://heasarc.gsfc.nasa.gov/xanadu/xspec/manual/XSappendixStatistics.html}}.
 
The outer PWN was used as the background for the inner PWN because there is some apparent structure in the inner PWN, which is embedded in the larger scale PWN. Thus, some emission from the large scale PWN may contaminate the spectra from these structures. To test the impact of the background used, we also fit the smaller-scale PWN using an offset, source free background region but found that the fitted spectral parameters agreed within the 1$\sigma$ uncertainties of those using the outer PWN as the background. However, the uncertainties are larger when using the outer PWN as the background,  so we conservatively report  the spectral parameters from these fits. 

 In the new XMM-Newton observation, the MOS1 and MOS2 data were relatively unaffected by background particle flares and these detectors collected about as many counts as collected in total from all three EPIC detectors during the first observation (ObsID 0802930101) reported by \cite{2019ApJ...875..107H}. Therefore, we jointly fit the spectra from the new and old observations to place the tightest constraints possible on the source spectrum. For the new observation, we extracted the source  spectra from the MOS1 and MOS2 Full Frame images using a 25$''$ radius circle centered on the source position, while the background spectra were extracted from a $\sim65''$ radius circle placed in a source free region. For the previous observations, we re-extracted the source spectra so that the  latest calibration and version of SAS were used for both data sets to limit any systematic offsets that these differences could introduce. The spectra were extracted from the Full Frame MOS1, MOS2, and PN event lists using a $25''$ circular region while the background was taken from  a $\sim86''$  source free circular region. The XMM-Newton spectra were binned to have 50 counts per bin prior to fitting, so that $\chi^2$ statistics could be used.

 All spectral fits in this paper were performed using XSPEC version 12.11.1 \citep{1996ASPC..101...17A}. We fit the spectra from each observation jointly, but multiply the models by a constant (freezing the constant factor to 1 for ObsID 22455) to account for any changes in the detector response due to the continuing build up of a contaminating layer on the detector over the $\sim$6 month interval between observations\footnote{See \url{https://cxc.cfa.harvard.edu/ciao/why/acisqecontamN0010.html}}. Similarly, we add the same constant for the XMM-Newton spectra, but fix the constant factor to 1 for the PN data from the older observation (i.e., ObsID 0802930101). We used the T\"{u}bingen-Boulder interstellar absorption model (tbabs) with the solar abundances of \cite{2000ApJ...542..914W}. Uncertainties in this paper are all reported at the 1 $\sigma$ level for a single interesting parameter unless otherwise noted.

\section{Results}
\label{results}

\begin{figure*}
\centering
\includegraphics[trim={0 0 0 0},scale=0.51]{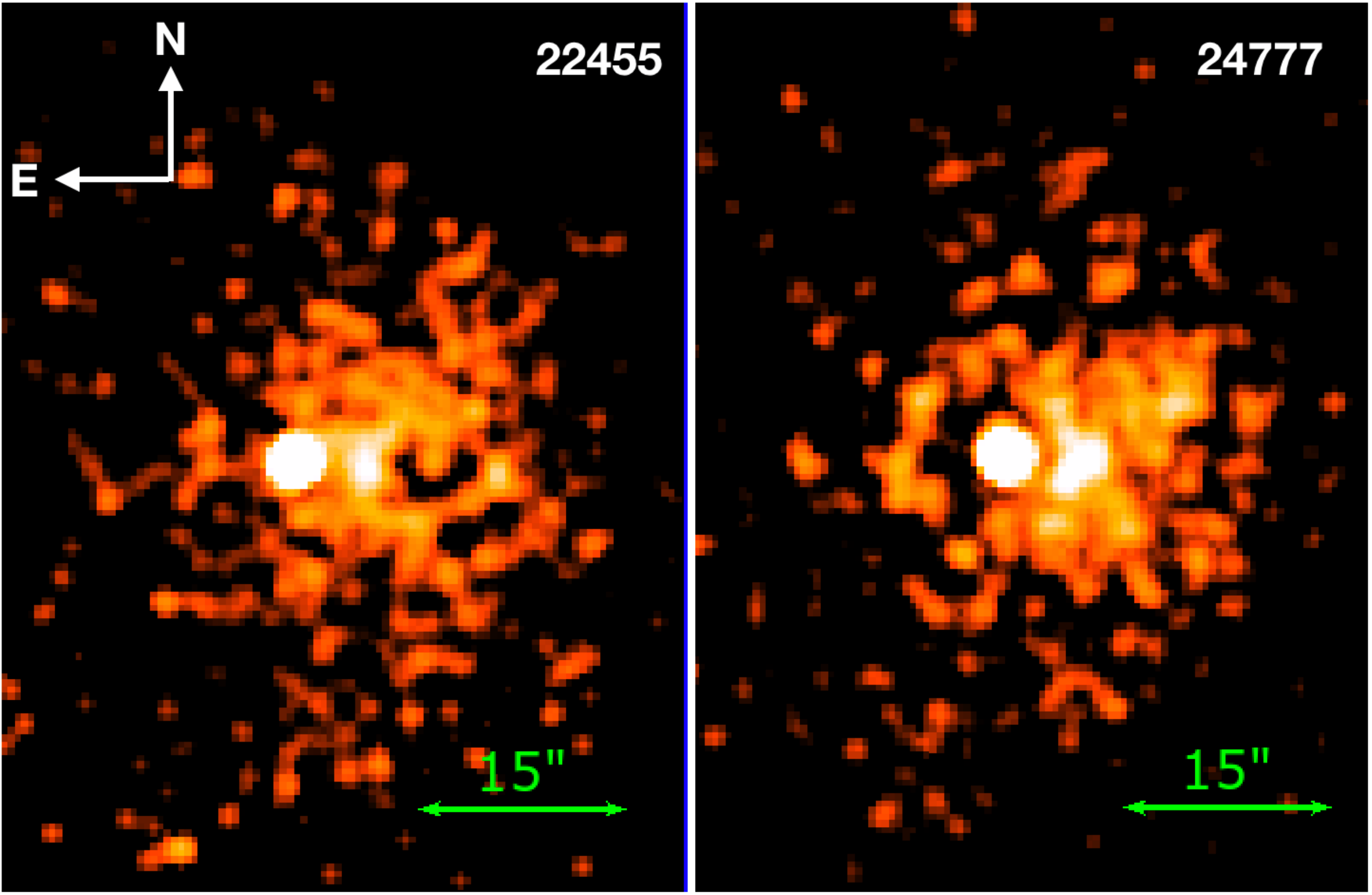}
\caption{CXO ACIS 0.5-8 keV exposure corrected images of the candidate pulsar and PWN J1015 from each individual ObsID. The images are smoothed with an $r=3''$ Gaussian kernel. The jet emission appears to change brightness between the two observations.} 
\label{expo_innerPWN}
\end{figure*}

\subsection{X-ray variability and timing}

\cite{2019ApJ...875..107H} previously searched the XMM-Newton Full Frame mode data, including MOS1, MOS2, and PN, for a period in the 10$^{-4}-6.8$ Hz frequency range, but found no statistically significant signal. However, given that most young pulsars with bright PWNe have pulse periods on the order of 10s to 100s of milliseconds \citep{2008AIPC..983..171K}, and the limited time resolution that Full Frame mode provides (i.e., 73.4 ms and 2.6 s for PN and MOS, respectively) this is unsurprising. In the new observation the PN detector was operated in Timing mode, offering a time resolution of 0.03 ms and allowing us to search for shorter periodic signals in the data. We extracted events in three energy bands (i.e., 0.5-10 keV, 0.5-2 keV, and 2-10 keV) from the detector columns between (and including) 34 and 42, which is where the source clearly dominates over the background. We used the $Z_m^{2}$ test with $m=1$ to search for the period \citep{1983A&A...128..245B}. The data spanned a total of $\Delta T=49.4$ ks, corresponding to a native frequency resolution of $\Delta\nu=(\Delta T)^{-1}=2\times10^{-5}$ Hz.  
In our search we used a factor of 3 higher frequency resolution, but for the largest peaks we performed a narrow search using an oversampling factor of 10. The largest peak found by \cite{2019ApJ...875..107H} at a frequency of $\sim5.87$ Hz had 
only a $2.6 \sigma$ significance. We started our search of the new pn timing mode data at 0.1 Hz and extended it up to a frequency of 100 Hz, corresponding to a 10 ms period, which is 
smaller than the
typical  periods of young pulsars with PWNe.

The largest peak found in the power-spectrum, 
calculated from $\sim 17450$ total counts with $\sim 4830$ counts from the point source, is located at a frequency of 35.314867(2) Hz with $Z_1^{2}=40.2$. We conservatively estimate the number of independent frequencies, or trials, as  $(\nu_{\rm high}-\nu_{\rm low})/\Delta\nu\approx4.9\times10^{6}$. After accounting for trials, this peak has only a $\approx2.8\sigma$ significance (see Figure \ref{timing_510_keV}). We also split the  data into soft (0.5-2 keV) and  hard (2-10 keV) energy bands to search for pulsations, but did not find any additional peaks with higher or comparable significance. In both energy bands the 35.3 Hz peak is still dominant, but it has a lower power and  remains  insignificant. For pulsars with complex pulse shapes, a significant peak of $Z_m^{2}$ can occur for higher harmonics, so we also calculated  $Z_m^{2}$ with $1<m<5$ across the entire frequency range (in the 0.5-10 keV energy range) and used the H-test to assess significance (see e.g., \citealt{2010A&A...517L...9D}). No other statistically significant peaks were found.

Assuming that the pulsations are nearly sinusoidal we use two methods to estimate the upper limit on the intrinsic pulsed fraction of the source. The first method follows the analytical approach of \cite{1975ApJS...29..285G}. 
The highest $Z^{2}_1=40.2$ value found in our search corresponds to the power defined in \cite{1975ApJS...29..285G}, $P \equiv Z_1^2/2 =20.1$. 
We derive a $3\sigma$ upper limit on the source power of $P_s\sim40$ (corresponding to a $Z^{2}_1\approx80$) using Figure 1 in \cite{1975ApJS...29..285G}. This can be converted into an observed pulsed fraction $p$ by using $Z^{2}_{1}=(N_{\rm tot}p^{2}/2)+2$ (see e.g., equations C5-C7 in \citealt{2021ApJ...923..249H}), where $N_{\rm tot}=17450$ is the total number of counts, 
which gives a 3$\sigma$ upper limit of $p<0.095$. This can then be converted into an intrinsic pulsed fraction by multiplying by $N_{\rm tot}/N_{\rm s}\approx3.61$, where $N_{\rm s}$ is the number of source counts, which gives $p_{\rm int}<0.34$, or a root-mean-square (RMS) intrinsic pulsed fraction, $p_{\rm int,RMS}=p/\sqrt{2} <0.24$ (see e.g., \citealt{2021ApJ...923..249H} for various commonly used pulsed fraction definitions and how they are related).

The second method to estimate the 3$\sigma$ upper-limit on the intrinsic pulsed fraction is using simulations. We first estimated the XMM background count rate by averaging the number of counts over detector columns with no source contribution. Next, we estimated the XMM count rates from the point source and extended emission by using the best fit spectral models from CXO, where the source is resolved. We then normalized the count rates from the point source and extended emission such that we get the total count rate when added to the averaged background count rate. The estimated XMM point source count rate was found to be 0.103 cts s$^{-1}$, while the background count rate was 0.27 cts s$^{-1}$, where the PWN contributes 0.135 cts s$^{-1}$ to the background.  We then used these count rates to simulate 200 observations of a sinusoidal signal, having the same total duration as the XMM Timing mode observation, while increasing the intrinsic RMS pulsed fraction from $8\%$ up to 35$\%$. These simulations show that intrinsic RMS pulsed fractions larger than about 20$\%$ would be detected at the 3$\sigma$ level, while pulsed fractions of about 25$\%$ would be detected at the 5$\sigma$ level. Therefore, we adopt the more conservative 3$\sigma$ upper-limit on the intrinsic pulsed fraction of $p<34\%$ (or $p_{\rm int,RMS}<24\%$). 

\subsection{X-ray spectra}
\label{xspectro}

We first fit the new XMM-Newton MOS1 and MOS2 data jointly with the pn, MOS1, and MOS2 data from the previous observation to place the tightest constraints possible on the source model parameters. Similar to \cite{2019ApJ...875..107H}, we find that the spectra are well fit by an absorbed BB+PL. The best-fit model parameters are 
listed in the first line of Table \ref{spec_fits} and the fitted spectrum and residuals are shown in Figure \ref{full_xmm_spec}. The obtained values agree with those previously found within their $1\sigma-2\sigma$ uncertainties.

Next, we fit all of the spectra from the CXO data. In all CXO spectral fits the absorbing column density is poorly constrained but consistent with the value found with XMM-Newton. Therefore, in all CXO fits we simply freeze the absorbing column density to the value obtained from the joint XMM-Newton spectral fits (i.e., $N_{\rm H}=2.8\times10^{21}$ cm$^{-2}$). The X-ray spectrum of the  point source, resolved by CXO, was previously well fit by an absorbed BB model \citep{2019ApJ...875..107H} so this is the model we start with for the new fits. This fit gives a slightly higher temperature ($kT=0.205\pm0.009$ keV) than the one found by \cite{2019ApJ...875..107H}, and a C-stat$=232.2$ for 211 degrees of freedom. However, large residuals above $\sim3$ keV suggest that an additional model component is necessary. Adding a PL component provides a statistically better fit to the data with C-stat$=202.5$ for 2 fewer degrees of freedom. The best fit parameters and their uncertainties are listed in Table \ref{spec_fits}, while the fitted spectrum is shown in Figure \ref{full_xmm_spec}. 
Unfortunately, there are few counts above 3 keV, leading to poor constraints on the photon index of the PL component. Given that the point source is embedded in an extended PWN, we also checked that the excess higher energy emission, modeled as a PL, is not due to undersubtracting the PWN emission. To accomplish this we chose the brightest part of the surrounding PWN (i.e., the magenta region in Figure \ref{expo_innerPWN}) 
as the background spectrum. However, even with this conservative background choice, the background counts still remain $<3\%$ of the total counts and the PL component is still required.

\begin{figure*}
\centering
\includegraphics[trim={0 0 0 0},scale=0.51]{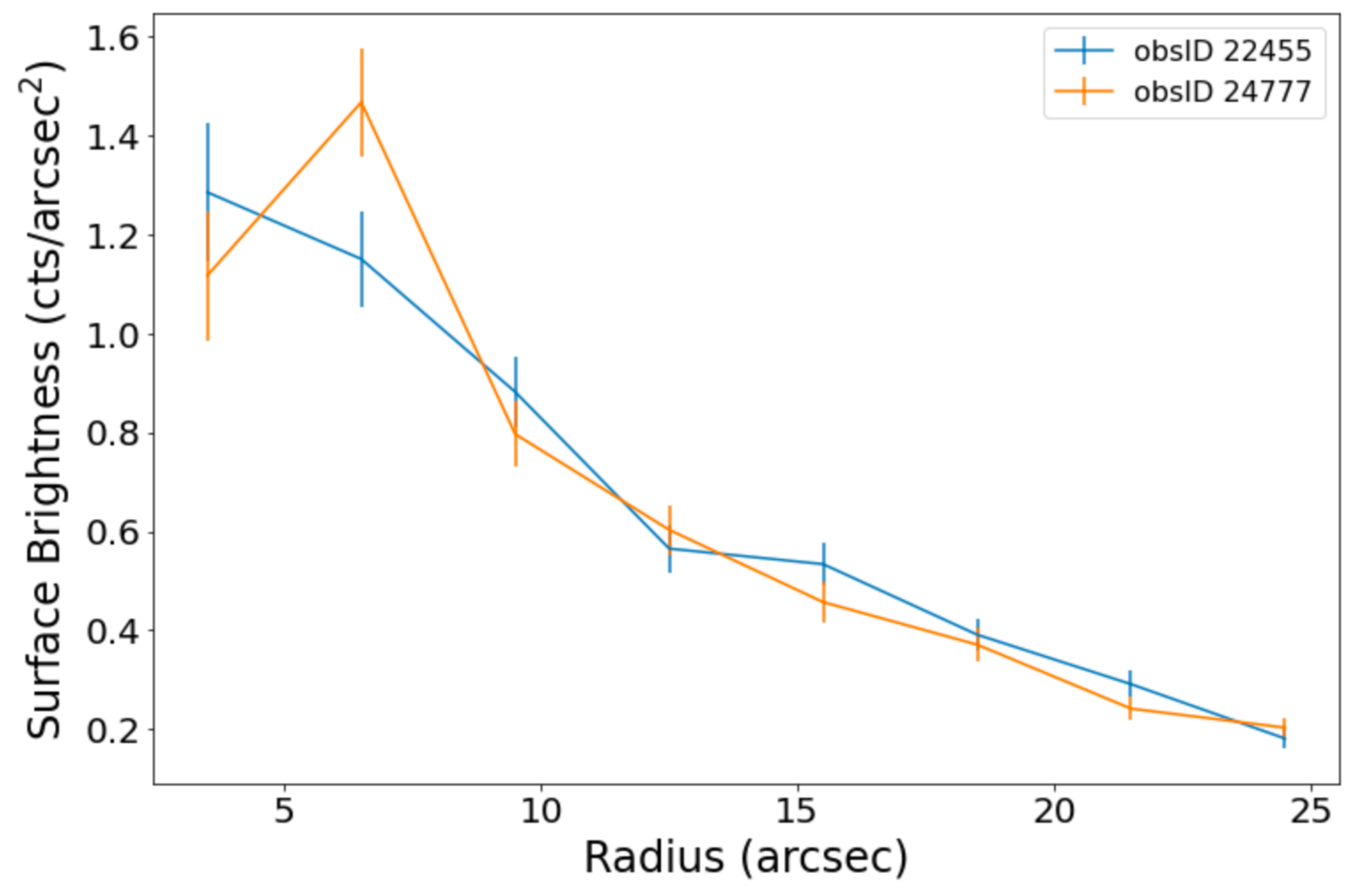}
\caption{Surface brightness of the extended emission versus radius from the point source for both observations. The ``jet'' like emission is marginally brighter in ObsID 24777.}
\label{radial_prof}
\end{figure*}

\begin{figure*}
\centering
\includegraphics[trim={0 0 0 0},scale=0.25]{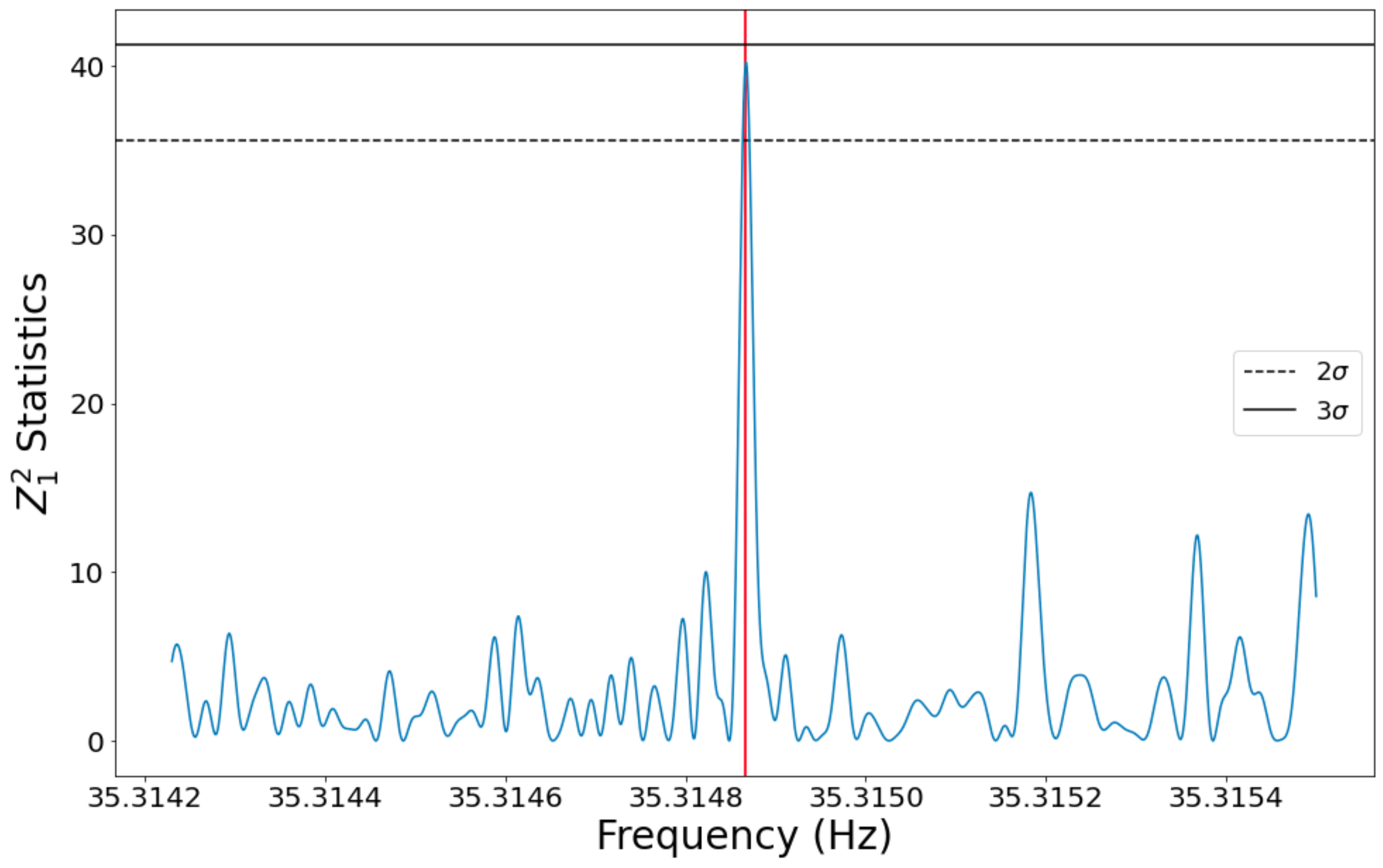}
\includegraphics[trim={0 0 0 0},scale=0.272]{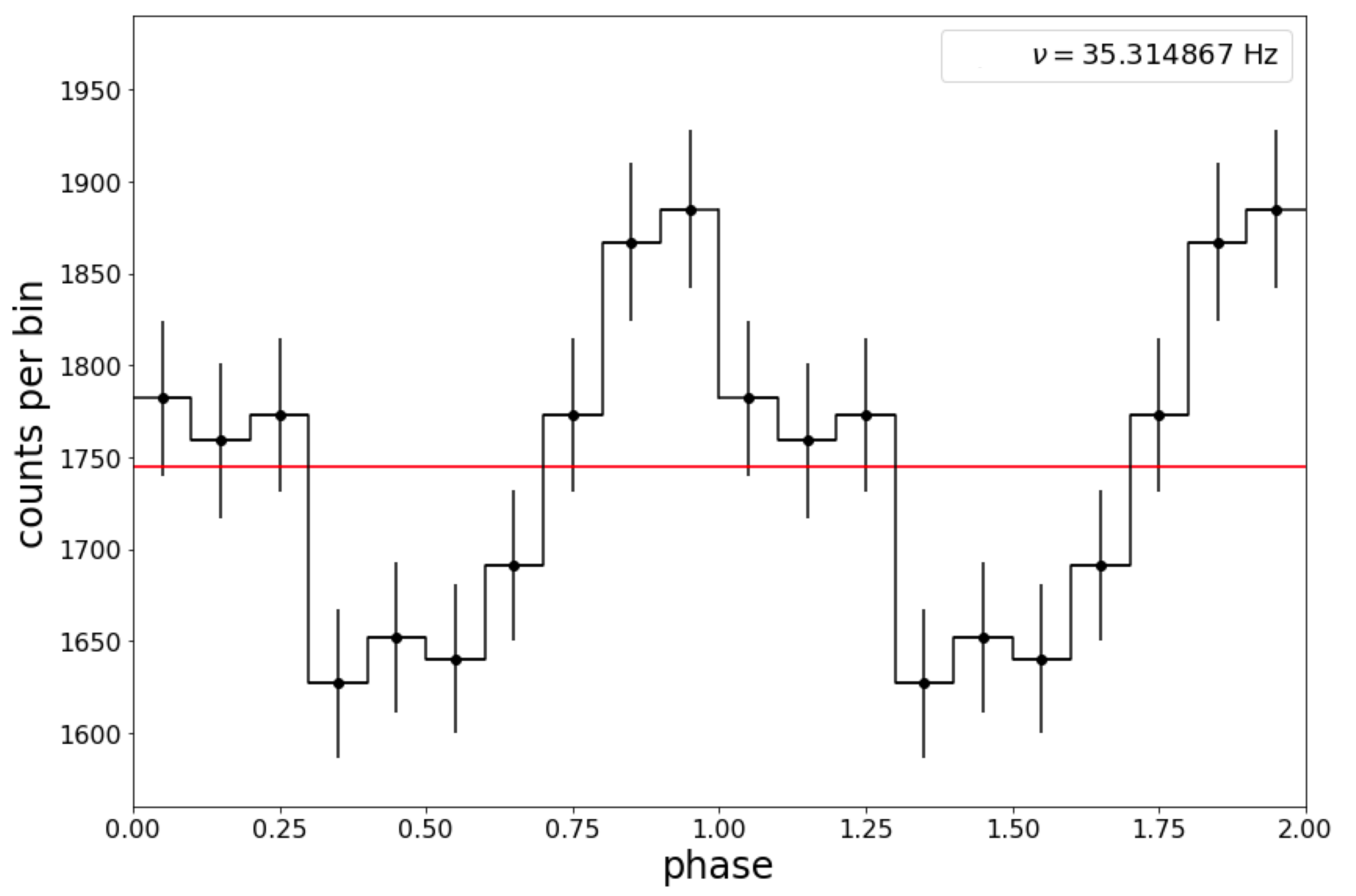}
\caption{{\sl Left:} $Z_1^{2}$ versus frequency near the largest peak found at 35.314867(2) Hz. The black dashed and solid lines correspond to the 2$\sigma$ and 3$\sigma$ significance $Z_1^{2}$ values. {\sl Right:} XMM-Newton pn Timing mode data folded on the 35.314867(2) Hz period. The red line shows the mean number of counts per bin. These pulsations are detected at a significance of  2.8$\sigma$, thus are not significant.
\label{timing_510_keV}
}
\end{figure*}

\begin{figure*}
\centering
\includegraphics[trim={0 0 0 0},scale=0.3]{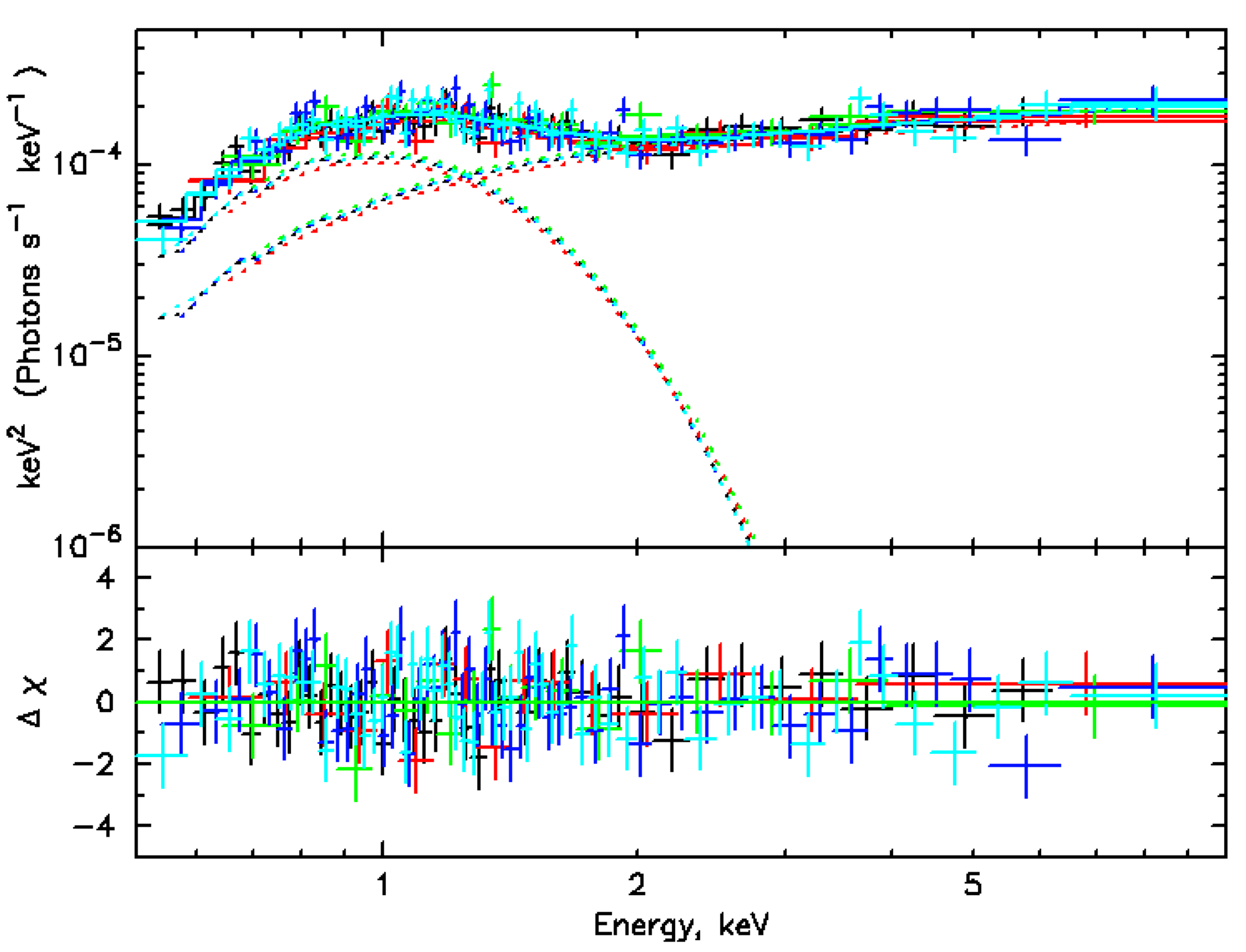}
\includegraphics[trim={0 0 0 0},scale=0.298]{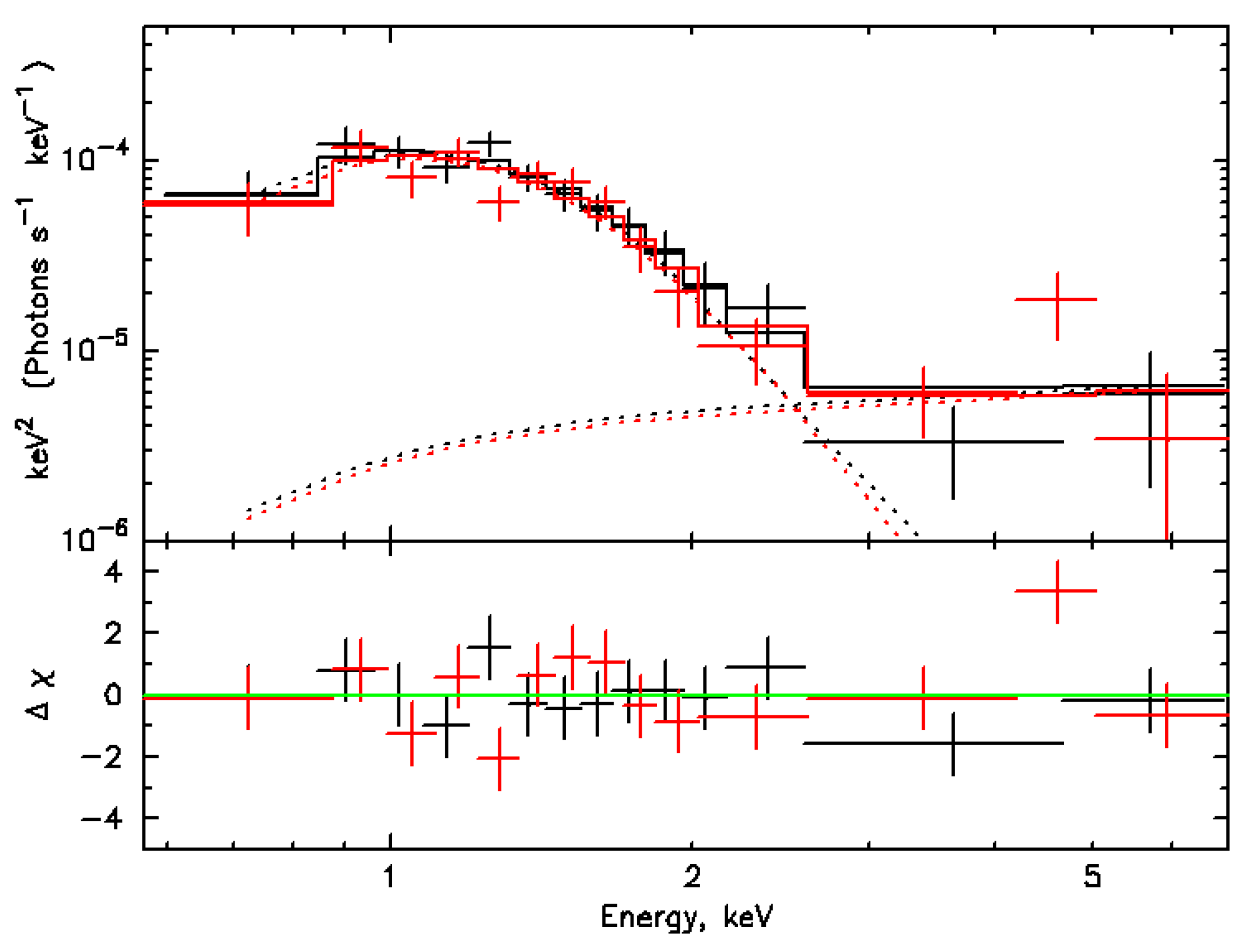}
\caption{{\sl Left:} The unfolded $\nu F_{\nu}$ XMM-Newton pn, MOS1, MOS2 (in red, black, and green, respectively)  spectra and MOS1, MOS2 (in blue and cyan, respectively) spectra from both old and new observations (i.e., OBSIDs 0802930101 0853180101, respectively), jointly fit with a blackbody$+$power-law model. XMM-Newton is unable to resolve the PWN from the point source. The bottom panel shows the residuals. {\sl Right:} The unfolded $\nu F_{\nu}$ CXO spectra (ObsIDs 22455 and 24777 are shown in red and black, respectively) of the point source fit with a blackbody$+$power-law model. The bottom panel shows the residuals. The Chandra spectra were binned for visualization purposes.
\label{full_xmm_spec}
}
\end{figure*}

We then fit the spectra of the full, small-scale inner, and outer PWN emission with an absorbed power-law model. The full PWN emission is well fit with $\Gamma=1.70\pm0.05$ and a similar flux in both observations (see Table \ref{spec_fits}). Similarly, the outer PWN's spectra are well fit by a PL with a comparable photon index $\Gamma=1.74\pm0.07$ to that of the full PWN. The small-scale PWN, however, shows a $\sim30\%$  change increase in its flux between the two most recent CXO observations (the change in the number of counts is $\sim10\%$). This, coupled with the change of radial profile observed in the small scale PWN between observations (see Figure \ref{radial_prof}), suggests that there may be some difference in the spectra between observations. Therefore, we fit the small-scale PWN spectra from each observation separately with an absorbed power-law model to search for differences in their photon index and normalization. Unfortunately, we lack the statistics to measure the photon indices and normalizations accurately enough to claim a  difference between the two epochs in terms of spectral model parameters. Therefore, we fit the spectra jointly to improve the overall statistics and find a best fit $\Gamma=1.55\pm0.12$ (see Table \ref{spec_fits}), which is harder the the spectrum of the full and outer PWN.

\begin{table*}[h]
\caption{Best fit models to the XMM-Newton and CXO spectra of the point source and extended emission.} 
\label{spec_fits}
\begin{center}
\renewcommand{\tabcolsep}{0.11cm}
\begin{tabular}{lcccccccc}
\tableline 
Region &  Const. & $F^{\rm unabs}_{\rm 0.5-10 keV}$ &  $N_{\rm H}$ &	$K_{\rm PL}$\tablenotemark{h} &	$\Gamma$ & $R_{\rm BB}$\tablenotemark{i} &  $kT$ & C-stat/dof \\
\tableline 
 & & 10$^{-13}$ c.g.s. & 10$^{21}$ cm$^{-2}$ &	10$^{-6}$ & 	& {m}  & { eV} \\
 \tableline 
 PS+PWN$_{\rm XMM}$\tablenotemark{a} &0.94(4)/1.05(4)/1.01(3)/1.01(3)\tablenotemark{c} & 9.8$\pm0.2$ & 2.8$^{+0.4}_{-0.3}$ & 105$^{+11}_{-10}$ & 1.69$\pm0.07$ &  1170$^{+260}_{-170}$ & 190$\pm10$ & 176.1/190\tablenotemark{g}\\
PS$_{\rm CXO}$\tablenotemark{b} &0.94$^{+0.08}_{-0.07}$\tablenotemark{d} & 3.3$\pm0.3$ & 2.8\tablenotemark{e} & 4.5$^{+11.0}_{-3.4}$ & 1.8$^{+0.9}_{-1.0}$ &  940$^{+150}_{-110}$ & 205$\pm9$ & 202.5/209\\
Full PWN$_{\rm CXO}$\tablenotemark{b} & 1.00$\pm0.05$\tablenotemark{d} & 7.4$\pm0.3$ &  2.8\tablenotemark{e} &  117$\pm6$ & 1.70$\pm0.05$ & ... & ... & 610.9/661 \\
Outer PWN$_{\rm CXO}$\tablenotemark{b} & 0.92$\pm0.06$\tablenotemark{d} & 4.1$\pm0.2$ & 2.8\tablenotemark{e} & 67$\pm5$ & 1.74$\pm0.07$ & ... & ... & 439.6/518\\
Inner PWN$_{\rm CXO}$\tablenotemark{b} & 1.32$^{+0.17}_{-0.15}$\tablenotemark{d} & 1.5$\pm0.2$ & 2.8\tablenotemark{e}  & 20$\pm3$ & 1.55$\pm0.12$ & ... & ... & 363.1/379\\
Inner PWN$_{22455}$\tablenotemark{f} & ... & 1.6$\pm0.2$ & 2.8\tablenotemark{e} & 18$^{+4}_{-3}$ & 1.40$\pm0.18$ & ... & ... & 173.9/186\\
Inner PWN$_{24777}$\tablenotemark{f} & ... & 2.0$\pm0.2$ & 2.8\tablenotemark{e} & 30$^{+5}_{-4}$ & 1.66$\pm0.16$ & ... & ... & 188.0/192 \\
\tableline 
\end{tabular} 
\tablenotetext{a}{Spectra of point source (PS) and pulsar wind nebula (PWN) were fit simultaneously from both XMM observations.}
\tablenotetext{b}{Spectra were fit simultaneously from both CXO observations.}
\tablenotetext{c}{Constant values of MOS1 and MOS2 from the 2017 and 2020 spectra (i.e, M1$_{17}$, M2$_{17}$ and M1$_{20}$, M2$_{20}$, respectively) of the source with respect to the pn spectrum from 2017.}
\tablenotetext{d}{Constant value from ObsID 24777 with respect to ObsID 22455.}
\tablenotetext{e}{Frozen}
\tablenotetext{f}{Parameters from individually fit spectra.}
\tablenotetext{g}{The $\chi^2$/dof is reported for this fit.}
\tablenotetext{h}{Power-law normalization in photons keV$^{-1}$ cm$^{-2}$ s$^{-1}$ at 1 keV.}
\tablenotetext{i}{ Blackbody radius defined as $R_{\rm BB}=(d/{\rm 10\, kpc})K_{BB}^{1/2}\,\,{\rm km}$, where $K_{BB}^{1/2}$ is the fitted model normalization and assuming a distance of 2 kpc.}
\end{center}
\end{table*}

\section{Discussion}
\label{discuss}

\subsection{Physical properties of the point source}

It is difficult to constrain the physical parameters of the putative pulsar (e.g., characteristic age, spin-down energy, magnetic field) given that no pulsations have been detected at X-ray energies. We found that the 3$\sigma$ upper-limit on the intrinsic pulsed fraction is about 34$\%$. Similar instances of putative pulsars surrounded by extended X-ray emission and lacking pulsations are known (see e.g., IC 443; \citealt{2015ApJ...808...84S}), while in other pulsars the intrinsic pulsed fraction can be lower than the upper-limit for J1015 (see e.g., Table 1 in \citealt{2020MNRAS.492.1025C}). However, we can still use the spectral properties of the point source to place some constraints on the physical parameters of the putative pulsar. For instance, the best fit $N_{\rm H}=2.8\times10^{21}$ cm$^{-2}$ can be converted to $E(B-V)\approx0.41$ using the relationship of \cite{2009MNRAS.400.2050G} corresponding to a rough distance estimate of  $d\approx2$ kpc using the {\tt mwdust} 3D dust maps \citep{2016ApJ...818..130B}. Taking the uncertainties of the $N_{\rm H}$ into account we find the range of distances are between approximately 1.8-3.1 kpc. We adopt a 2 kpc distance for the remainder of the discussion, but note that this method of distance estimation is prone to additional large uncertainties (e.g., due to the use of different $E(B-V)$ maps, $N_{\rm H}$ to $E(B-V)$ relationships, ISM abundances).

The spectral parameters derived from the BB fits of the CXO spectrum of the point source are in agreement with that found by XMM-Newton and reported by \cite{2019ApJ...875..107H}, but CXO uncovered an additional PL component, likely coming from the magnetosphere of the pulsar. The unabsorbed flux from the PL component ($F_{\rm 0.5-10 \ keV}=2.6^{+2.1}_{-0.7}\times10^{-14}$ erg cm$^{-2}$ s$^{-1}$) is fainter than that from the BB component ($F_{\rm 0.5-10 \ keV}=3.1^{+0.3}_{-0.2}\times10^{-13}$ erg cm$^{-2}$ s$^{-1}$) by about an order of magnitude. These fluxes correspond to unabsorbed luminosities of $L_X=4\pi d^2F_{(\rm 0.5-10 \ keV)}=1.2\times10^{31}d_2^2$ erg s$^{-1}$ and $L_X=1.5\times10^{32}$ erg s$^{-1}d_2^2$, where $d_2=d/$(2 kpc). We find an emitting radius of the BB component $R_{\rm BB}=(d/{\rm 10\, kpc})K_{BB}^{1/2}\,\,{\rm km}=940^{+150}_{-110}d_2$ m and corresponding bolometric thermal luminosity $L_{\rm bol}\approx2\times10^{32}d_2^2$ erg s$^{-1}$.

Thermal emission often dominates the non-thermal X-ray emission in middle-aged gamma-ray pulsars with characteristic ages $\tau_c\gtrsim 100$ kyr, such as Geminga, PSR J0659+1414, PSR B1055-52, and PSR J1740+1000 \citep{2014ApJ...793...88M,2018ApJ...869...97A,2010ApJ...720.1635M,ApJsub...B1055,2022MNRAS.513.3113R}. These rotation powered pulsars, having very similar $\dot{E}$ values, $\sim10^{34}-10^{35}$ erg s$^{-1}$, typically show two BB components, always with a hotter component coming from a smaller emitting region, and a cooler component coming from a larger emitting region (see e.g., \citealt{2007ApJ...670..655K,2014ApJ...793...88M,2016ApJ...833..253K}). The obtained BB radius is comparable to those of hot polar caps observed in most of the above-mentioned pulsars. No cold BB component is observed in J1015, likely due to the relatively high $N_H$ and large distance to the source compared to the above-mentioned pulsars. 

The  temperature  of $\sim0.2$ keV measured for J1015 is at the upper end of the 0.1-0.2 keV interval of the hotter components found in other middle-aged pulsars. However, J1015's bolometric thermal luminosity and PL slope are similar to  middle-aged pulsars, which typically have $L_{\rm bol}\approx10^{31}-10^{32.5}$ erg s$^{-1}$ and $\Gamma=1.6-1.8$ (see e.g., \citealt{2013MNRAS.434..123V,2008AIPC..983..171K}). Based on these similarities, we assume that J1015 belongs to this group of middle-aged gamma-ray pulsars. Middle-aged pulsars often have  X-ray efficiencies of $\eta_X \equiv L_X/\dot{E} =10^{-3}-10^{-4}$ \citep{2008AIPC..983..171K}, suggesting that J1015 may have an $\dot{E}\sim(10^{34}-10^{35})d_2^2$ erg s$^{-1}$.

There is no plausible counterpart to the source in the relatively deep Pan-STARRs optical survey \citep{2019ApJ...875..107H}. While isolated pulsars are not expected to be detected by ground-based optical surveys, it is possible that the source may have a binary companion. Therefore, we also searched for a counterpart in the more sensitive Dark Energy Camera Plane Survey 2 (DECaPS2; \citealt{2018ApJS..234...39S,2023ApJS..264...28S}), which reaches a $g-$band limiting magnitude of 24.0, and a $y-$band limiting magnitude of 21.2. The closest DECaPS2 counterpart is located $1.4''$ away from J1015, which is well outside of the 3$\sigma$ positional uncertainty of the source thus ruling out any optical counterpart to the source. We also searched for a counterpart to the jet-like feature to ensure it is not emission from a chance coincident background/foreground source, but did not find any counterpart. Furthermore, J1015 has also not been detected at radio energies by Parkes (private communication S. Johnston). Lastly, J1015 is fainter in flux at GeV energies than all but one young radio-quiet pulsar, PSR J0622+3749\footnote{PSR J0622+37489 was discovered in a blind search of the Fermi-LAT data \citep{2012ApJ...744..105P}}, reported in the Fermi 2nd pulsar catalog \citep{2013ApJS..208...17A}. Of the X-ray detected radio quiet gamma-ray pulsars, J1015 has a relatively small GeV to non-thermal X-ray flux ratio $F_{\rm GeV}/F_{\rm X, nonth}\approx 680$, with only the 290.4 ms pulsar PSR J1958+2846 having a lower $F_{\rm GeV}/F_{\rm X, nonth}\approx 670$\footnote{We note that \cite{2012ApJS..201...37K} report a factor of 10 smaller X-ray flux than that used by \cite{2013ApJS..208...17A} to calculate $F_{\rm GeV}/F_{\rm X, nonth}$. }. However, there are several young radio loud X-ray detected GeV pulsars that have lower ratios.

\subsection{Physical properties of the extended emission}

The morphology of the extended emission around the point source resembles those of some other PWNe around middle-aged rotation-powered pulsars (see e.g., \citealt{2006ARA&A..44...17G,2008AIPC..983..171K,2017SSRv..207..175R} for reviews). On large angular scales, the PWN has a rounded shape (possibly resembling a bow-shock), while the inner region shows a hint of a ring-like torus surrounding the pulsar (particularly visible in ObsID 24777) with a bright knot to the  west of the pulsar that may be emission from a jet-like outflow. There is a slight hint of variability in the jet-like emission, but deeper observations are required to confirm this variability (see Section \ref{img_analysis}). The bow-shock shape of the PWN may suggest that the pulsar has escaped its host supernova remnant and is now travelling in the ISM, where ambient sound speeds are much lower. This would suggest an age larger than a few tens of kiloyears. The photon index ($\Gamma=1.7$) of the PWNe's spectrum and the ratio of the PWN to point source flux are largely consistent with other known PWNe/PSRs \citep{2008AIPC..983..171K}. 
At the assumed distance of $\sim2$ kpc, the PWN has an unabsorbed PWN luminosity is $L_{\rm 0.5-10 \  keV}\approx4\times10^{32}$ erg cm$^{-2}$ s$^{-1}$. The efficiency of PWNe (i.e., $\eta=L_X/\dot{E}$) is generally $\eta<10^{-1}$ suggesting that the spin-down energy of this pulsar should be  $\dot{E}>4\times10^{33}$ erg s$^{-1}$ in agreement with those of  GeV pulsars, which have $\dot{E}\gtrsim10^{34}$ erg s$^{-1}$ \citep{2008AIPC..983..171K,2013ApJS..208...17A}. Overall, it seems that the inferred properties of the J1015 pulsar and PWN indicate that it is likely a middle-aged pulsar with an age of a few 100 kyr and spin-down luminosity $\do{E}\simeq10^{34} - 10^{35}$ erg s$^{-1}$. 

The observed X-ray morphology of the PWN is very much reminiscent of that of the PWN powered by the young ($\tau\approx10$ kyrs) and energetic ($\dot{E}=2.2\times 10^{37}$ ergs s$^{-1}$) PSR J2229+6114 \citep{2001ApJ...552L.125H} (see panel \#6 in Figure 2 of \citealt{2008AIPC..983..171K}), which is also detected in GeV \citep{2009ApJ...706.1331A}. This pulsar is characterized by a very low X-ray radiative efficiency of the PWN, $\eta_{\rm X, pwn}=4\times10^{-5}$ (at the typically assumed distance of 3 kpc). However, the non-thermal flux from PSR J2229+6114 (whose spectrum also exhibits a thermal component)  is comparable to the PWN flux, unlike J1015 whose non-thermal magnetospheric component is much fainter than the PWN. The GeV efficiency of PSR J2229+6114 is about $10^{-2}$ and $F_{\rm GeV}/F_{\rm X, nonth}\approx 380$ \citep{2008AIPC..983..171K,2009ApJ...706.1331A}.  The PWN of J1015 would also resemble the Vela PWN if the latter was placed at a larger distance and was more absorbed, such that only the bright compact part of the PWN was discernible. Note that Vela PWN is has fairly low radiative efficiency, $\eta_{\rm X, pwn}=1\times10^{-4}$, but $F_{\rm GeV}/F_{\rm X, nonth}\approx 2700$.

\section{Summary and Conclusion}
\label{sumandconc}

In conclusion, the new X-ray observations of J1015 have allowed us to better constrain the properties of the point source and extended emission surrounding it. We find that the point source's spectrum is well fit by an absorbed BB+PL model with $kT=0.205\pm0.009$ keV, $R_{\rm BB}=940^{+150}_{-110}$ m (assuming a 2 kpc distance) and $\Gamma=1.8^{+0.9}_{-1.0}$, which is consistent with thermal emission from a hotspot on the surface of a NS and magnetospheric component that dominates at higher X-ray energies. No pulsations are observed from the source, with a 3$\sigma$ upper limit on the intrinsic pulsed fraction of 34$\%$.  

The extended emission is resolved and shows several features, including a possible torus, jet, a bow-shock like morphology, and a hint ($\approx2.2\sigma$) of variability between the two CXO observations. However, additional deeper X-ray observations are needed to confirm the variability. The extended emission's spectrum is well fit by an absorbed PL model with $\Gamma=1.70\pm0.05$. All of these features are consistent with other PWNe.

The X-ray properties of the J1015 point source and extended emission strongly suggest that it is a radio-quiet middle-aged pulsar (i.e., $\tau\gtrsim100$ kyr), similar, e.g., to the well studied Geminga pulsar. X-ray observations with a high angular and timing resolution observatory could also allow for a detection of pulsations at lower pulsed fractions. Deeper radio observations could also potentially be used to detect pulsations from the source, but it may be that the radio beams are misaligned with Earth as is the case for many radio quiet GeV pulsars. Lastly, blind searches at GeV energies may also be able to detect pulsations (see e.g., \citealt{2012ApJ...744..105P}), but if the pulsar is noisy or shows glitches, this approach my be difficult/unsuccessful.

\medskip\noindent{\bf Acknowledgments:}
 We thank the anonymous referee for providing useful feedback which helped to improve the clarity and readability of the manuscript.
Support for this work was provided by the National
Aeronautics and Space Administration through Chandra Award GO0-21064A issued by the Chandra X-ray Observatory Center, which
is operated by the Smithsonian Astrophysical Observatory for and on behalf of the National Aeronautics and
Space Administration under contract NAS8-03060.
JH acknowledges support from an appointment to the NASA Postdoctoral Program at the Goddard Space Flight Center, administered by ORAU. JH thanks Brad Cenko for helpful discussions related to the access and use of DECam data and Simon Johnston for providing information about radio observations of the source.

\end{document}